\documentclass{kluwer}    
\usepackage{graphicx}

\begin{document}

\newcommand{\acena}{\mbox{$\alpha$~Cen~A}}
\newcommand{\ms}{\mbox{m\,s$^{-1}$}}
\newcommand{\muHz}{\mbox{$\mu$Hz}}

\begin{article}
\begin{opening}         
\title{Oscillations in $\alpha$ Cen A observed with UCLES at the AAT} 
\author{T.~R. \surname{Bedding}$^1$}
\author{R.~P. \surname{Butler}$^2$, C. \surname{McCarthy}$^2$}  
\author{H. \surname{Kjeldsen}$^3$}
\author{G.~W. \surname{Marcy}$^4$}
\author{S.~J. \surname{O'Toole}$^2$}
\author{C.~G. \surname{Tinney}$^5$}
\author{J. \surname{Wright}$^4$}

\institute{$^1$School of Physics, University of Sydney 2006, Australia\\
  $^2$Carnegie Institution of Washington, Department of
  Terrestrial Magnetism, \\\quad 5241 Broad Branch Road NW, Washington, DC
  20015-1305\\
  $^3$Theoretical Astrophysics Center, Aarhus University,
  DK-8000, Aarhus C, \\\quad Denmark\\
  $^4$Department of Astronomy, University of California, Berkeley, CA
  94720 USA\\
  $^5$Anglo-Australian Observatory, P.O.\,Box 296, Epping, NSW 1710,
  Australia}

\runningauthor{Butler et al.}
\runningtitle{Oscillations in $\alpha$ Cen A}

\begin{abstract}
We report Doppler measurements of $\alpha$~Cen~A from time-series
spectroscopy made with UCLES at the 3.9-m AAT.  Wavelength calibration
using an iodine absorption cell produced high-precision velocity
measurements, whose power spectrum shows the clear signature of solar-like
oscillations, confirming the detection reported by Bouchy \& Carrier
(2001).

\end{abstract}

\end{opening}           

\section{Observations and Reductions}  

We were awarded six nights to observe \acena{} in May 2001 with UCLES and
the iodine cell at the 3.9-m Anglo-Australian Telescope (AAT).  The run was
affected by bad weather (the first night was completely lost), with an
overall usability of about 50\%.  We obtained 5169 spectra of \acena, with
one spectrum every 20\,s.  We were also awarded four nights at the VLT to
use UVES with the iodine cell.  Again, only 50\% of the time was usable.

Data processing was postponed by our development of a new raw reduction
package (which will also be used for the planet-search program).  The
revised software properly treats the echelle blaze function, removes about
90\% of the cosmic rays, and includes a telluric filter.  The net result is
an improvement of about 0.5 m/s in the velocity precision of each spectrum.
Velocities have been extracted from the UCLES/AAT spectra, but processing
of the UVES/VLT data is not yet complete.

\section{Velocities}

\begin{figure}
\centerline{\includegraphics[width=\the\hsize]{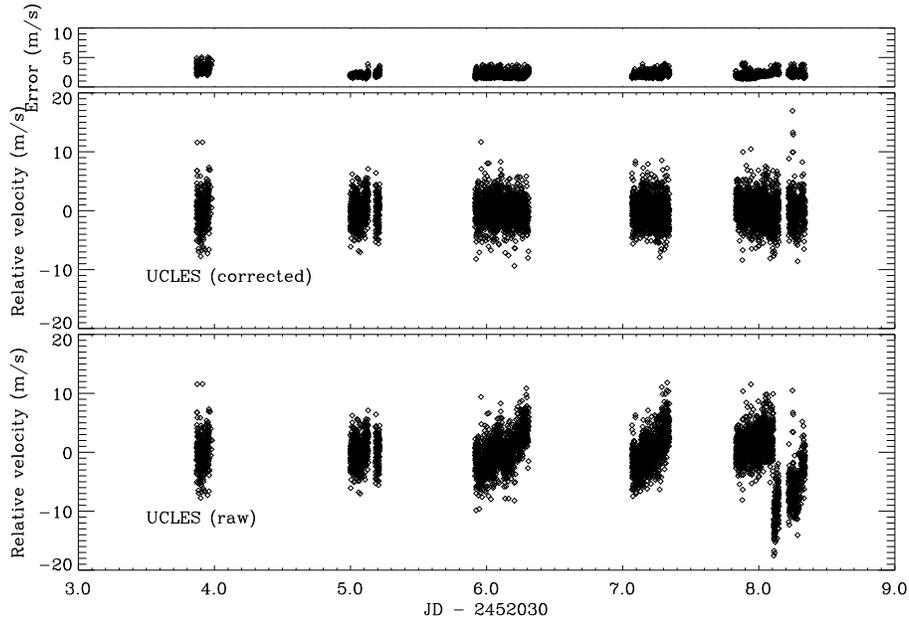}}
\caption{Time series of velocity measurements of \acena.}
\label{fig.vel}
\end{figure}

The raw UCLES velocities (bottom panel of Fig.~\ref{fig.vel}) show slow
trends during night 3--5.  The trends are related to the slow movement of
the dewar as liquid nitrogen boils off.  Nights 3 and (particularly) 5 also
show a jump that coincides with the refilling of the CCD dewar.

Note that these observations differ from our earlier run on $\beta$~Hyi
\cite{BBK01} in that the CCD was rotated by 90 degrees to speed up readout
time.  This also forced the readout shift to be in the direction of
dispersion, and has clearly introduced systematic drifts.

We have corrected for the drifts and jumps by fitting and subtracting a
mean trend from the affected segments of the time series.  This is
effectively a high-pass filter, the effect of which can be seen in the
middle panel of Fig.~\ref{fig.vel}.  Once this is done, the rms velocity
precision at high frequencies is 2.2\,\ms{} per spectrum.

\clearpage

\section{Analysis}

\begin{figure}
\centerline{\includegraphics[width=\the\hsize]{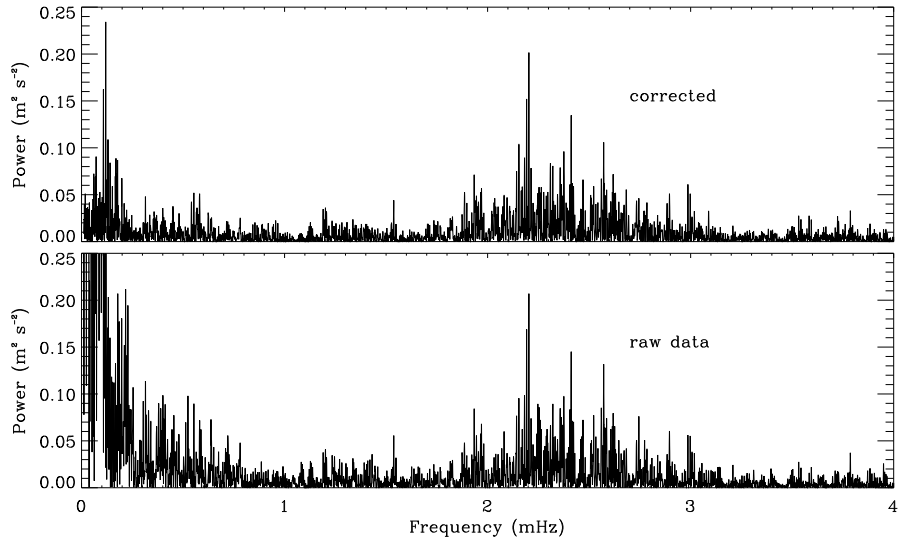}}
\caption{Power spectrum of \acena, showing an excess of power around
  1.9--3.0\,mHz.}
\label{fig.acen}
\end{figure}

The power spectra of the raw and corrected time series are in
Fig.~\ref{fig.acen}.  The high-pass filtering greatly reduces the noise at
low frequencies, as expected.  The power excess agrees well with that seen
from CORALIE observations by \inlinecite{BC2001}.

\begin{figure}
\centerline{\includegraphics[width=\the\hsize]{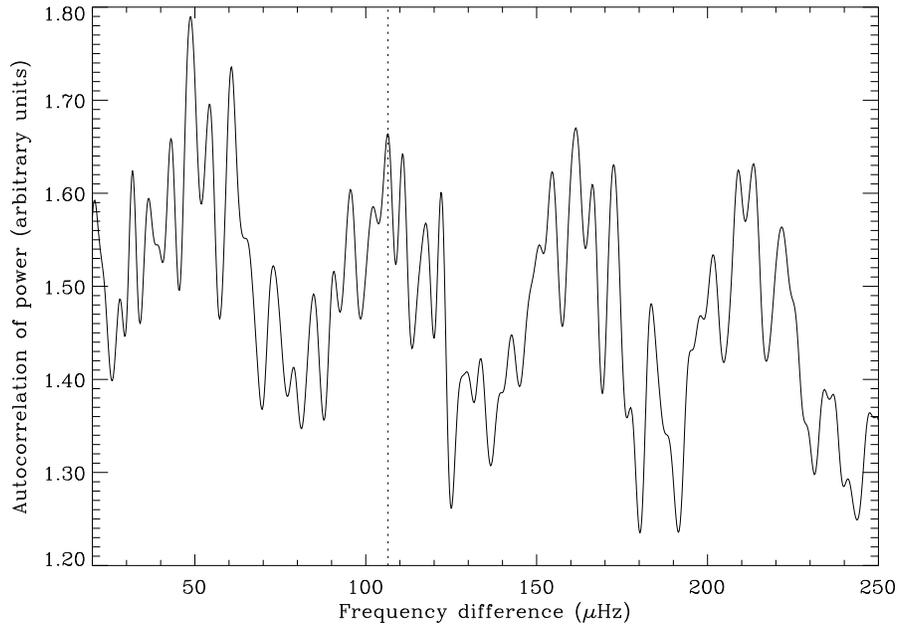}}
\caption{Autocorrelation of the power spectrum.  The dotted line is at
 106.5\,\muHz.}
\label{fig.auto}
\end{figure}

\begin{figure}
\centerline{\includegraphics[width=\the\hsize]{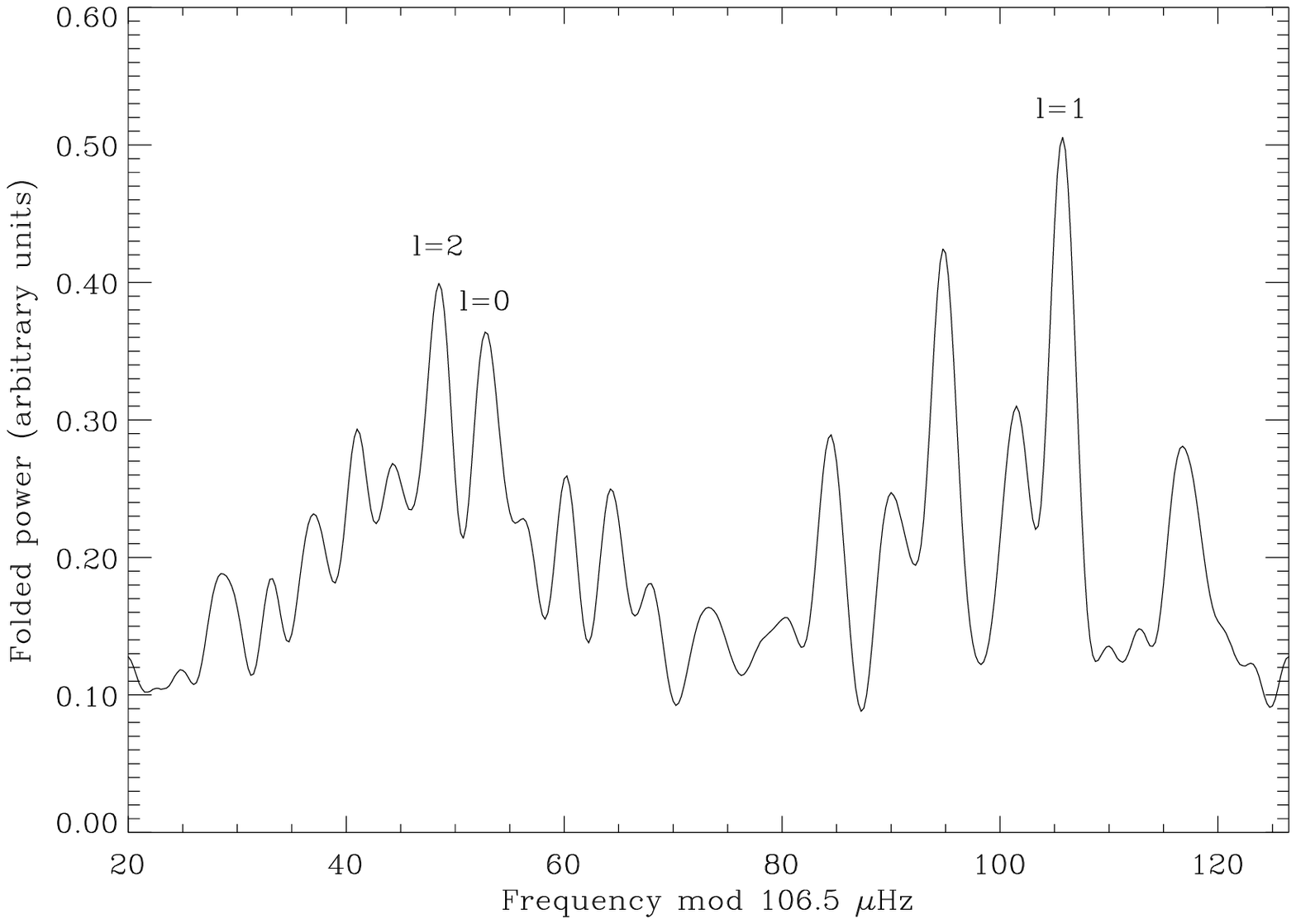}}
\caption{Folded power spectrum of \acena.  Mode identifications by
\protect\inlinecite{BC2002} are shown.}
\label{fig.folded}
\end{figure}

Figure~\ref{fig.auto} shows the autocorrelation of the power spectrum in
the region 1.7--3.1\,mHz.  The pattern is consistent with p-mode
oscillations having a large separation of $\Delta\nu \simeq 106.5$\,\muHz.
The folded power spectrum in the region 1.7--3.1\,mHz is shown in
Figure~\ref{fig.folded}.  Mode identifications by \inlinecite{BC2002} are
indicated.

Once the UVES/VLT observations are reduced, the spectral window will be
greatly improved and we hope to be able to measure individual mode
frequencies for $l=0, 1, 2$ and perhaps~$3$.

\end{article}
\end{document}